\documentclass[numbered]{trbunofficial}

\usepackage[hidelinks]{hyperref}
\usepackage{subcaption}
\usepackage{float}
\usepackage{graphicx}
\usepackage{subcaption}  
\usepackage{xcolor}  
\usepackage[shortcuts]{extdash}
\usepackage{graphicx}
\usepackage{url}
\usepackage{color, soul}
\usepackage{bm}
\usepackage{amsmath} 
\usepackage{lineno}
\usepackage{siunitx}
\title{
    Investigating Resiliency of Transportation Network Under Targeted and Potential Climate Change Disruptions
}

\AuthorHeaders{Rahimitouranposhti, Sharma, Camur, Omitaomu, and Li}

\author{
  \textbf{Maedeh Rahimitouranposhti}\\
  mrahimi1@vols.utk.edu\\
  Department of Industrial and Systems Engineering, University of Tennessee, Knoxville, TN 37996, USA \\
  \hfill\break% this is a way to add line numbering on empty line
  \textbf{Bharat Sharma, Ph.D.} \\
  sharmabd@ornl.gov\\
  Computational Sciences and Engineering Division, Oak Ridge National Laboratory, Oak Ridge, TN 37830, USA \\
  \hfill\break%
  \textbf{Mustafa Can Camur, Ph.D.} \\
  mcamur@utk.edu\\
  Department of Industrial and Systems Engineering, University of Tennessee, Knoxville, TN 37996, USA \\
    \hfill\break%
  \textbf{Olufemi A. Omitaomu, Ph.D.} \\
  omitaomuoa@ornl.gov\\
  Computational Sciences and Engineering Division, Oak Ridge National Laboratory, Oak Ridge, TN 37830, USA \\
    \hfill\break%
  \textbf{Xueping Li, Ph.D.} \\
  xueping.li@utk.edu\\
  Department of Industrial and Systems Engineering, University of Tennessee, Knoxville, TN 37996, USA 
}
\begin{document}
\maketitle
\section{Abstract}
Ensuring robustness and resilience in intermodal transportation systems is essential for the continuity and reliability of global logistics. 
These systems are vulnerable to various disruptions, including natural disasters and technical failures.
Despite significant research on freight transportation resilience, investigating the robustness of the system after targeted and climate-change driven disruption remains a crucial challenge.
Drawing on network science methodologies, this study models the interdependencies within the rail and water transport networks and simulates different disruption scenarios to evaluate system responses. 
We use the data from the US Department of Energy Volpe Center for network topology and tonnage projections.
The proposed framework quantifies deliberate, stochastic, and climate‑driven infrastructure failure, using higher resolution downscaled multiple Earth System Models' simulations from Coupled Model Intercomparison Project Phase version 6.
We show that the disruptions of a few nodes could have a larger impact on the total tonnage of freight transport than on network topology. 
For example, the removal of targeted 20 nodes can bring the total tonnage carrying capacity to 30$\%$ with about 75$\%$ of the rail freight network intact. 
This research advances the theoretical understanding of transportation resilience and provides practical applications for infrastructure managers and policymakers. 
By implementing these strategies, stakeholders and policymakers can better prepare for and respond to unexpected disruptions, ensuring sustained operational efficiency in the transportation networks.

\hfill\break%
\noindent\textit{Keywords}: Transportation Resilience, Freight Transport System, Climate Change Disruptions, Targeted Disruptions, Robustness Analysis
%\newpage

\section{Introduction}
\label{Introduction}

 In an increasingly interconnected world, the resilience of transportation systems has become a critical area of focus for research and development. 
 Intermodal transportation presents unique challenges and opportunities in ensuring robustness against various disruptions, including natural disasters, technical failures, and security threats \cite{Warner}. Recent studies have sought to address these challenges by developing advanced models and frameworks that enhance system resilience's assessment and improvement. Resilience in intermodal transportation refers to the system's ability to anticipate, absorb, adapt, and quickly recover from disruptions. It ensures that transportation networks, such as those combining rail and truck services, can maintain functionality and efficiency even when faced with unexpected events like natural disasters or man-made attacks \cite{schofer2022resilience}.

 We synthesize insights from several recent papers that explore resilience in different segments of transportation networks, including rail-truck freight networks, maritime supply chains, and airport networks. Each study contributes to a multi-dimensional understanding of resilience by introducing novel analytical frameworks, optimization models, and strategic management practices tailored to these networks' unique characteristics and vulnerabilities. These multifaceted approaches highlight the complexities of maintaining resilience in intermodal transportation systems and underscore the need for robust and flexible strategies to manage and mitigate the significant negative impacts of disruptions. 

 Disruption in transportation systems refers to events that cause significant negative impacts on transportation system operations.
 These disruptions can arise from natural disasters and climate extremes (e.g., earthquakes, floods, hurricanes), man-made events (e.g., cyber, terrorist attacks, strikes), or other unexpected incidents that compromise the infrastructure, decrease capacity, or interrupt the normal flow of goods and services \cite{wan2018resilience}.
 Disruptions challenge the system's ability to function effectively, requiring robust and flexible strategies to manage and mitigate their effects.

 A key aspect of ensuring resilience is the robustness of the intermodal transportation network, which is critical for maintaining functional performance despite disruptions. Robustness in intermodal transportation refers to the ability of the transportation network to retain its functional performance despite disruptions or perturbations \cite{he2021robustness}. It measures how well the network can avoid direct and indirect economic losses and continue operating when facing disruptions to transport elements like infrastructure failures or capacity reduction.

 This study comprehensively analyzes rail and water networks within the  United States (US), which are critical components of global freight transportation, by evaluating their resilience under various disruption scenarios, including climate extreme events \cite{Sharma_2022_IEEE} and targeted disruptions(e.g., cyberattacks). The study includes a comprehensive literature review of system disruption, robustness, and resiliency, modeling these networks, simulating disruption scenarios, and evaluating the effectiveness of various resilience measures, particularly emphasizing the impact of rising temperatures crossing critical thresholds. The goal is to provide actionable insights for infrastructure managers and policymakers to enhance the resilience of intermodal transportation systems.

 Hereon, the paper is organized as follows. We begin with a literature review that categorizes and analyzes studies on resilience in transportation systems. We then explore the robustness and resilience of US rail and water networks under various disruption scenarios, including the effects of rising temperatures and targeted disruptions. Our analysis includes examining the impact of identifying critical network nodes. Finally, we discuss the broader implications for enhancing system resilience, offering recommendations for infrastructure managers and policymakers, and suggesting future research directions.

\section{Literature Review}
\label{Literature Review}

In this section, we examine the existing literature on resilience within intermodal transportation systems, encompassing road, rail, water, and air components. Our analysis categorizes and dissects various types of disruptions that impact these modes, focusing on strategies for effectively managing and mitigating such disruptions. By synthesizing key insights from recent studies, this review highlights the multifaceted approaches necessary to enhance the resilience of intermodal transportation networks.

Resilience in transportation systems has been extensively studied over the years; however, the specific resilience of intermodal systems has received comparatively limited attention \cite{murray2006comparison}. Resilience in this context is evaluated by the system’s performance in response to unexpected disruptions. A tentative measure of resilience is proposed as the percentage reduction in performance indices of the intermodal transportation system following such disruptions \cite{zhang2009framework}. This measure underscores the importance of not just immediate recovery but also the system's capacity to adapt and sustain functionality under stress.

A significant focus of this review is on the natural disaster of heat waves, which present a critical disruption to transportation networks. With the predicted increase in the number of hot days per year, key nodes within intermodal networks are at heightened risk of recurring heat waves. This makes the analysis of such disruptions essential for future resilience planning. Previous studies, such as \citep{schofer2022resilience}, have largely emphasized the immediate response and short-term resilience of specific modes, like the rail industry, during crises. However, our analysis extends this discussion by highlighting that heat wave disruptions are not merely transient events; they are likely to persist and intensify with time. This underscores the necessity for long-term resilience strategies, including sustained adaptations and the implementation of permanent infrastructural and operational changes to mitigate the ongoing and worsening impacts of climate-related disruptions.

\subsection{Transportation Systems Resilience}

The network characteristics vary by mode of transportation and are driven by factors such as nature and space for infrastructure. The quantitative analysis of the resilience of airport networks is presented by \citep{clark2018resilience}, where the authors utilized network science to assess robustness against disruptions like natural disasters, technical failures, and man-made threats. The paper investigates the US airport network where nodes are airports and edges are bi-directional flights. It highlights the importance of network attributes in maintaining critical functions and developing effective recovery strategies. Notably, the study finds that disruptions of only 30$\%$ of airports can cause a complete collapse of the airport network. 
The paper not only evaluates the cascading impacts of disruptions but also identifies optimal recovery strategies. Specifically, the results show that betweenness centrality (a measure of the importance of nodes within the network) plays a critical role in determining the most effective recovery order for both partial and full recovery phases. This insight underscores the value of prioritizing key nodes to enhance recovery efficiency, offering valuable guidance for proactive resilience planning and infrastructure design.

Extending the network science approach to rail networks, \citep{bhatia2015network} provide a comprehensive network science-based framework to measure and compare the resilience of the Indian Railways Network against various hazards such as human-induced and natural. 
The authors utilize simulations inspired by real-world events, namely the 2004 Indian Ocean Tsunami and the 2012 North Indian blackout, to illustrate the network's hazard responses and recovery strategies. The study employs multiple metrics, including degree, strength, and centrality measures, to generate various recovery sequences. Findings suggest that recovery strategies based on betweenness centrality are most effective, highlighting the importance of restoring key bridge stations to enhance the network's functionality rapidly. This quantitative framework is generalizable across large-scale critical lifeline infrastructure networks and can inform stakeholders on optimal recovery strategies for different types of hazards, ensuring improved resilience and preparedness.

Furthermore, \citep{schofer2022resilience} examine the US rail intermodal freight system's response to the COVID-19 pandemic. Their research builds on existing frameworks that explore the adaptability of supply chains under stress, emphasizing the critical role of intermodal freight in maintaining supply chain continuity during crises. 
%Previous studies have often focused on the resilience of individual transport modes or specific segments of the supply chain; however, 
This study integrates insights from both qualitative interviews and quantitative data to provide a comprehensive assessment of the rail industry’s performance during the pandemic. The authors also highlight the complexity of the intermodal freight system, involving multiple stakeholders and processes, which aligns with earlier research on the interconnectedness of global supply chains. By focusing on the pandemic as a unique stress test, this study adds to the growing understanding of how transport systems can be restructured to enhance resilience against future disruptions.
%%%

%%%%

In a related context, \citep{yadav2020resilience} present a hypothesis-driven resilience framework for urban transport networks, specifically applied to the London Rail Network (LRN). The study addresses the compounded vulnerabilities arising from natural disasters like floods and targeted cyber-physical attacks, emphasizing the critical inter-dependencies within urban transport networks. Utilizing network science principles, the authors demonstrate that the efficiency-focused design of the LRN makes it particularly susceptible to cascading failures during such compound disruptions. 
They find that the network's topology, optimized primarily for operational efficiency and cost minimization, lacks sufficient redundancy and robustness to withstand significant disruptions. The study concludes that enhancing resilience requires both pre-disruption planning and effective post-disruption recovery strategies, with network centrality-based recovery methods proving to be efficient. Importantly, the authors underscore the need for resilience-centric designs in urban infrastructure. This approach advocates for a paradigm shift where resilience is integrated into the planning and development of transport networks, moving beyond traditional design practices that focus solely on efficiency and structural longevity. Resilience-centric design involves incorporating redundancies, diversifying network pathways, and enhancing the flexibility of infrastructure to adapt to and recover from unexpected events. By prioritizing resilience in the design phase, urban transport networks can better mitigate the risks posed by increasing urbanization and climate change, reducing their vulnerability to both natural and man-made disruptions.
Shifting focus from rail to maritime transportation, \citep{young2020intermodal} provide an extensive overview of the complexities and vulnerabilities associated with intermodal maritime supply chains. The authors emphasize the significance of containerized shipping in global trade and highlight various security measures implemented post-9/11.

Despite these measures, vulnerabilities persist, particularly in the segments of the supply chain before goods reach US shores. The paper identifies several key factors that influence supply chain resilience, including the physical and informational flows of goods, the roles of various stakeholders, and the potential risks posed by cyber threats. 

%Addressing resilience against seismic disruptions, 

Following our analysis of general disruption types and mitigation strategies, it is essential to delve into specific frameworks that address resilience in intermodal networks under particular conditions. For instance, \citep{misra2022estimating} focus on enhancing resilience against seismic disruptions within intermodal freight networks. It introduces a probabilistic framework that integrates publicly available datasets and models to predict damage and manage recovery of network components. 
This framework is illustrated through a case study in Memphis, Tennessee, assessing the network's ability to maintain functionality and recover post-disruption. Key contributions include novel restoration models that directly link physical damage to network functionality, allowing for dynamic simulation of network performance and recovery. 

Another approach for evaluating seismic resilience is conducted by \citep{misra2019seismic}. The authors introduce a robust framework for assessing the seismic resilience of rail and truck intermodal networks, critical for long-haul freight transport in the US. They utilize a multi-scale modeling approach that considers local and national impacts on freight movement and employ Monte Carlo simulations to evaluate network performance under seismic disruptions. Key innovations include value-weighted connectivity and inverse travel distance metrics to assess network recovery and functionality post-disaster. The framework's application in a case study of the New Madrid seismic zone demonstrates its potential to provide detailed insights into economic impacts and infrastructure vulnerabilities.

Enhancing resilience in intermodal transportation networks is crucial for maintaining efficient and reliable supply chains. \citep{feng2024improving} explore strategies such as optimizing the use of empty containers and coordinating bulk cargo transport. It introduces a model that calculates the optimal number of empty containers to lease under uncertain demand, incorporating trip-sharing strategies to lower transportation costs and improve supply chain resilience.  
Complementing this, \citep{chen2012resilience} introduce a resilience indicator that quantifies the ability of the freight network to recover after various disruptions.
The author proposed a mixed-integer program to stochastically quantify a freight network's ability to recover after disruptions. The efficacy of the framework is demonstrated through experiments on a double-stack container network under various disaster scenarios, highlighting the critical role of network topology and recovery activities in enhancing resilience. The study concludes that network resilience is a multifaceted concept that involves both pre-disaster planning and post-disaster actions, providing a comprehensive approach to disaster management in intermodal freight transport systems.

Network‑science studies demonstrate that freight systems can suffer rapid functional collapse under three failure archetypes \cite{yadav2020resilience}.  The first is a targeted attack, in which an informed adversary deliberately disables the most central hubs—an effect modelled by descending centrality removal and illustrated by recent cyber or physical breaches at major ports \cite{clark2018resilience}. The second is a stochastic outage, where equipment failures, storm closures, or staffing shortages strike nodes at random \cite{aar2025economic}. We operationalise these two archetypes in our disruption scenarios by (i) removing nodes in descending centrality order and (ii) deleting nodes uniformly at random. A third, climate‑driven archetype is motivated by empirical heat‑wave research.  Field evidence shows that when daily maximum temperatures exceed approximately 35\si{\celsius} rail buckling, speed restrictions, and power‑supply failures become common \cite{ferranti2018hottest, clark2019vulnerability}.  Accordingly, our heat‑wave scenario removes any node whose projected annual count of hot days (T\textsubscript{max}>35\si{\celsius}) under SSP5‑8.5 (2021–2050) exceeds its 1991–2020 baseline \cite{wang2022future}, thereby representing recurrent, climate‑induced service suspensions (see Section \ref{Impact of Rising Hot Days} for more details).

\subsection{Disruption in Freight Transportation System}
This section examines  disruptions impacting intermodal freight transportation, focusing on the effects of extreme heat events, tropical cyclones, and other climate-related challenges. We explore the documented increases in global temperatures, as well as projections for future climate conditions, and analyze how these changes exacerbate vulnerabilities in transportation infrastructure. Through case studies from various cities, we detail the economic and operational impacts of these disruptions and review the resilience strategies proposed in the literature to mitigate their effects. The aim is to provide a comprehensive understanding of the challenges posed by climate extremes and the adaptive measures necessary to enhance the resilience of transportation systems.

To begin with heatwave disruption, \citep{kumar2021novel} address the impact of critical temperatures and heat waves on transportation infrastructure. It highlights that global mean surface temperatures have risen by 0.85$\si{\celsius}$ since the pre-industrial era, %with projections under the RCP4.5 scenario
indicating further increases of 0.3-0.7$\si{\celsius}$ short-term (2016-2035) and 1.1-2.6$\si{\celsius}$ long-term (2081-2100). These rising temperatures lead to road buckling and railway track deformation, disrupting transportation systems. The paper emphasizes the need for resilient infrastructure to withstand such impacts and proposes a framework incorporating flexibility, diversity, and industrial ecology to enhance infrastructure resilience. This includes strategies to adapt to increased frequency and intensity of heatwaves, ensuring the transportation sector can manage and mitigate these climate risks effectively.

Continuing with the theme of heatwave impacts, \citep{ferranti2018hottest} examine the impact of the record-breaking heatwave on July 1, 2015, where temperatures reached 37.5$\si{\celsius}$ at Heathrow Airport. This extreme heat caused rail tracks to expand and buckle, necessitating emergency speed restrictions to prevent derailments, while overhead lines sagged, leading to power loss for trains. Signaling systems, sensitive to heat, accounted for 57$\%$ of heat-related incidents, and telecommunications equipment overheated in line-side cabinets. Over 220,000 delay minutes were recorded, with significant delays on critical routes like London North Eastern, costing the national economy an estimated \pounds 16 million. The paper suggests resilience strategies, such as real-time monitoring with low-cost sensors and implementing green infrastructure, to adapt to the increasing frequency of heat waves projected for the future.

%%%%

%%%%
Expanding on the vulnerabilities of transportation infrastructure under extreme weather conditions, \citep{beheshtian2018impacts} address how critical temperatures and heat waves lead to transportation infrastructure failures. They emphasize the vulnerability of New York City's motor fueling infrastructure
during extreme weather events, such as Hurricane Sandy, approximately 67\% of gas stations in the New York metropolitan area were inoperable on the fourth day following the storm due to power outages and fuel supply disruptions. This percentage gradually decreased over time, reaching about 28\% by the eleventh day as stations regained functionality and fuel supplies were restored.  This widespread inoperability was primarily due to damaged refineries and terminals disrupting fuel supplies. They model the impact of these disruptions on fuel availability and travel behavior, finding that severe flooding events could render up to 85$\%$ of flood-vulnerable elements inoperable by the 2080s. Additionally, resilience-enhancing strategies like backup generators and fuel reservoirs are discussed to mitigate the impacts of extreme weather on transportation.

Transitioning to the specific impacts of extreme heat events (EHEs) significantly impacts transportation infrastructure in Phoenix, AZ, where summer temperatures average around 41$\si{\celsius}$ and can reach up to 50$\si{\celsius}$. These conditions cause asphalt roads to soften, leading to pavement rutting and reduced lifespan, and induce thermal expansion in steel bridge components, stressing joints and compromising structural integrity. High temperatures also result in engine overheating and increased tire blowouts in vehicles, while restricting construction activities due to safety concerns. Public transportation systems face disruptions from expanded steel rails and sagging overhead power lines. Notably, a 2011 incident in Mesa, AZ, where temperatures hit 41.7$\si{\celsius}$, triggered a transformer fire, causing power outages that affected over 100,000 homes and key infrastructure, illustrating the cascading failures extreme heat can provoke. Projections indicate a sixfold increase in EHE frequency and a doubling of event duration by 2070, necessitating adaptive strategies such as improved infrastructure design, increased vegetation, enhanced urban albedo, and comprehensive emergency response plans to mitigate these impacts and enhance resilience \cite{clark2019vulnerability}.

Further, \citep{feng2022tropical} investigate the risks of extreme heat events on infrastructure, including transportation, due to compound hazards from tropical cyclones and heatwaves. they highlight that heatwaves with a heat index over 40.6$\si{\celsius}$ significantly disrupt critical services. For instance, in Harris County, Texas, tropical cyclones like Hurricane Ike caused power outages affecting over 63$\%$  of residents for more than five days. Projections under the high-emissions scenario 
show that the probability of experiencing a heatwave lasting over five days following a cyclone will increase from 2.7$\%$ historically to 20.2$\%$ by the late 21st century. This will escalate the percentage of residents facing compound hazards from 0.8$\%$ to 18.2$\%$. The paper proposes resilience strategies such as undergrounding 5$\%$ of local power distribution networks, which could reduce the impact on residents experiencing extended outages and heatwaves from 18.2$\%$ to 11.3$\%$, underscoring the critical need for infrastructure adaptation to mitigate disruptions in transportation and other sectors.

In a similar vein, \citep{hatvani2018policy} discuss the impact of heatwaves on urban infrastructure, particularly focusing on transportation. 
During the severe heatwave in early 2009 in southern Australia, significant transport disruptions occurred in Adelaide and Melbourne. The rail systems were particularly affected, with heat-related failures leaving commuters stranded. They suggest several measures to mitigate these issues, such as cooling rails with cold water during heatwave alerts, replacing wooden rail sleepers with more heat-resistant concrete ones, and ensuring timely communication about non-operational train lines while providing alternative transportation like buses. they also note that in 2009-2010, 2$\%$ of South Australia's annual energy demand accounted for 65$\%$ of the yearly cumulative power generation costs, highlighting the critical need to manage peak electricity demand during heatwaves, which also impacts transportation systems dependent on reliable power supply.

Lastly, \citep{bolitho2017heat} examine the impact of extreme heat on infrastructure, particularly transportation systems, with detailed examples such as the 2009 heatwave in Melbourne. This event caused significant transportation disruptions, including train delays and failures due to rail track buckling and power supply issues. The study emphasizes the critical temperature threshold of 35$\si{\celsius}$, above which these failures become more frequent, predicting an increase from 9 to 12 days per year by 2030 and up to 27 days by 2100. It highlights the necessity for coordinated policy responses and cross-sectoral integration to mitigate these effects, suggesting improved infrastructure design, early warning systems, and community-based resilience initiatives to handle the increasing frequency and intensity of heat waves.

By synthesizing these studies, it becomes evident that the transportation sector must adopt comprehensive resilience strategies to mitigate the impacts of extreme weather events, particularly heat waves, on infrastructure.

\section{Methodology and Datasets}
\label{Analysis}

In this section, we describe the methodology and framework used to analyze the resilience and robustness of transportation networks. We begin by outlining the structure and modeling approach for the rail and water transportation systems, which form the basis of our analysis. The subsequent subsections delve into the centrality measures employed to assess the criticality of various nodes within the network, followed by an exploration of different disruption scenarios designed to evaluate the system's ability to withstand and recover from multiple types of failures. 

Daily maximum temperatures were obtained from the CMIP6 (Coupled Model Intercomparison Project Phase 6) Localized Constructed Analogs, version 2 (LOCA-2) archive \cite{LOCAref}, which statistically down‑scales each ESM to a 1/16‑degree grid (6 km) across North America. 
LOCA employs a constructed analog technique: for each ESM day, it identifies historical synoptic patterns that are dynamically similar, applies quantile mapping bias correction, and grafts the resulting local detail onto the future projection.
The method yields 6 km temperature fields that retain the parent model’s climate signal while adding the terrain‑driven nuance required for impact studies.

Finally, we introduce the metrics used to quantify the network's robustness and resilience, including assessments of how disruptions affect the network's operational capacity. This section provides a comprehensive overview of the analytical methods and models applied, setting the stage for interpreting the results.

Our objective is to provide a transparent, nation‑wide first‑pass assessment across more than 130 rail‑ and port‑nodes. Static network metrics—namely the largest connected component and its derivative State of Critical Functionality (SCF) \cite{bhatia2015network}—are computationally tractable at this scale and allow direct comparison of disruption archetypes and climate projections on a common footing. Because the model assumes no spontaneous rerouting once a node fails, the resulting capacity‑loss curves constitute an upper‑bound stress test: any real‑world adaptive response would shift outcomes upward, but not below this baseline. We recognise that dynamic queueing, surge capacity, and carrier‑specific rerouting are important; incorporating them through agent‑based or system‑dynamics extensions is identified in the Discussion as a priority for future work.

\subsection{Rail and Water Networks Data}The data used in this paper comes from the Freight and Fuel Transportation Optimization Tool (FTOT) developed by the US DOT Volpe Center \cite{volpe2023} to verify origin to destination for freight shipment connection. This tool is designed to optimize the routing and flow of commodities over multimodal transportation networks. It provides insights into the optimal cost, emissions, vehicle distance traveled, and facility utilization by mode and commodity. The FTOT can generate summary reports, maps, and visualizations and is equipped to facilitate scenario comparisons, including disruption and resilience analyses, making it highly suitable for studying the resilience of transportation systems under various disruptions. We model the US rail and water transportation system as an origin-destination network, with nodes represented by train stations and water ports.

FTOT is, to our knowledge, the most comprehensive public data source that uses the Freight Analysis Framework, Version 5 (FAF5) origin–destination matrices \citep{fhwa2022faf5}, Surface Transportation Board car‑load reports, U.S. Army Corps water‑borne statistics, and Class I railroad schedules into a single multimodal network. Even so, three limitations warrant mention. First, the tonnage projections inherit the macro‑economic assumptions and commodity aggregations embodied in FAF5. Second, facility locations are represented as centroid nodes, so fine‑scale details—such as individual port berths or rail spurs—are not resolved. Third, truck flows are inferred from commodity balances rather than measured directly, which can under‑represent niche, high‑value road freight. We therefore pair a purely structural indicator i.e., SCF with a tonnage‑weighted indicator to hedge against potential bias in any single FTOT value.

\begin{figure}
\begin{minipage}[b]{0.48\textwidth}
    \includegraphics[trim = {.1cm .1cm .0cm .7cm}, clip, width=\textwidth]{
        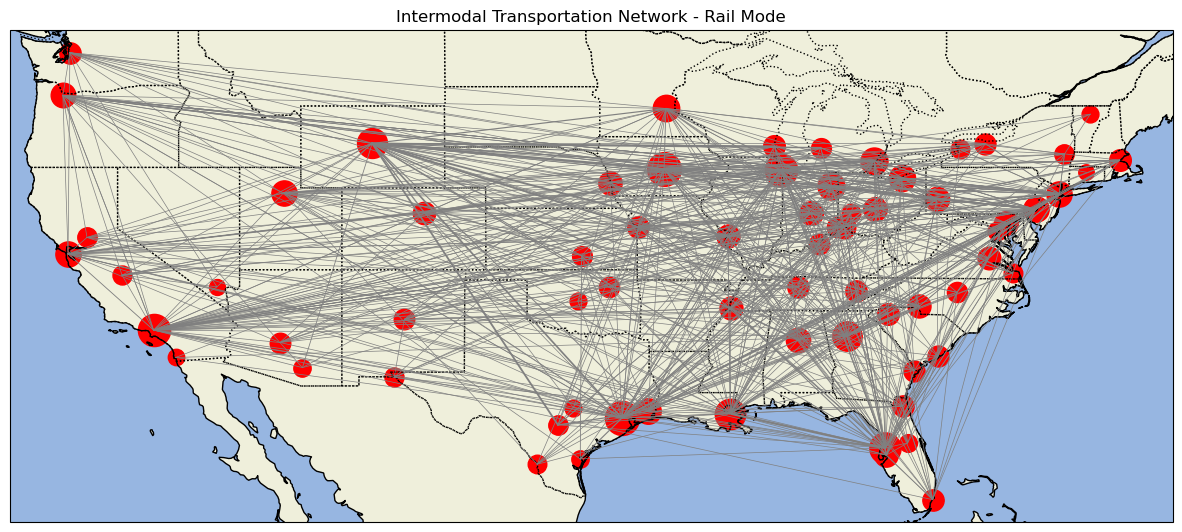}
    \caption*{(a) Rail Transport Network}
\end{minipage}
\hfill
\begin{minipage}[b]{0.48\textwidth}
    \includegraphics[trim = {.1cm .1cm .0cm .7cm}, clip, width=\textwidth]{
        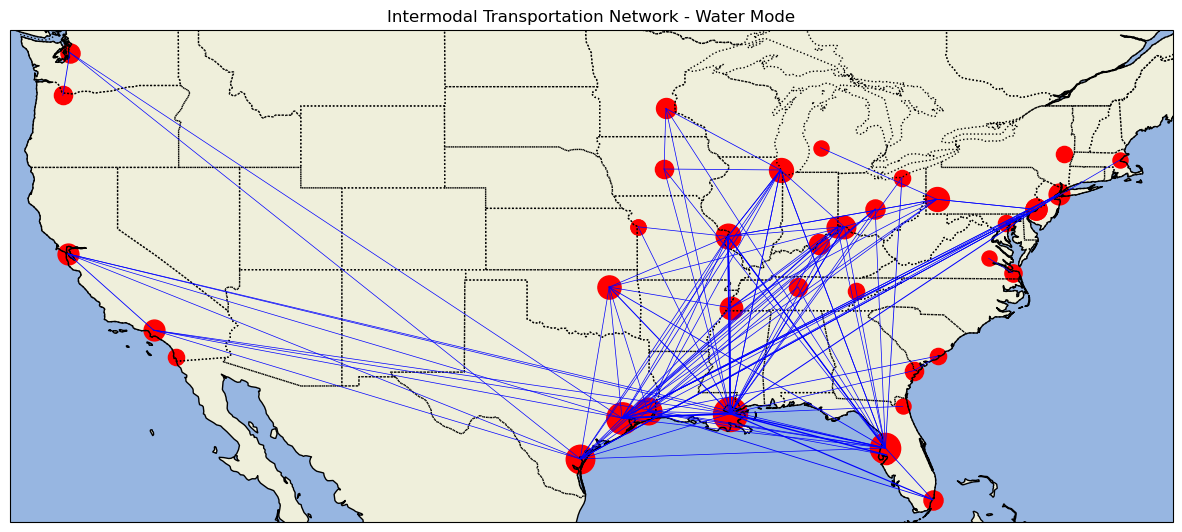}
    \caption*{(b) Water Transport Network}
\end{minipage}
\caption{\textbf{US Freight Network Topology Plots of Rail and Water Modes.}
Rail Transport Network (a) and Water Transport Network (b). 
Size of the nodes is proportional to the tonnage transported through them.}
\label{f:NetworkFig}
\end{figure}

The connectivity of the rail and water networks is illustrated in Figure~\ref{f:NetworkFig}. The rail transport network comprises 84 nodes with an average degree of 20.19, and the water transport network consists of 47 nodes with an average degree of 6.55. The average degree of a network is a measure of the connectivity of the nodes and is calculated by summing the degrees of all the nodes and then dividing by the total number of nodes in the network. In this case, the degree of a node refers to the number of direct connections (edges) it has with other nodes. A higher average degree indicates a more interconnected network, which can impact the network’s resilience and efficiency in responding to disruptions. In both subplots, node sizes are proportional to the total load (tonnage) transported from/through that respective node. All edges in this transport network are bi-directional, i.e., the movement could be back and forth from a node.

\subsubsection{Centrality Measures}

Understanding the significance of nodes, such as rail stations or water ports, within a transportation network is crucial for developing robust enterprise-level resilience strategies. Centrality, a key metric in network analysis, quantifies the importance of these nodes within the overall network structure \cite{wang2011exploring}. In this study, we apply centrality measures to quantify the extent of nodes within our network. Among the various available centrality measures, those relevant to our analysis include closeness and betweenness centrality. These measures for the rail and water networks were used to rank the top 10 nodes. 

Closeness centrality quantifies the extent to which a node (i) is `central' based on its proximity to all other nodes in the network and it is expressed as in Equation \eqref{closeness}.

\begin{equation} \label{closeness}
c_{CL}(i) = \frac{1}{\sum_{j \in V} dist(i, j)},
\end{equation}

where \(dist(i, j)\) represents the shortest network distance between nodes \(i\) and \(j\) in our network graph. For comparison with other centrality measures, \(c_{CL}\) is normalized to a range between \([0,1]\).

Betweenness centrality quantifies the degree to which a node acts as an intermediary between other pairs of nodes. Here, the `significance' of a node pertains to its strategic position within the network paths, which we represent as rail and water routes for freight movement. Nodes that lie on numerous routes are typically more vital to the overall network flow. For our calculations, we utilized betweenness centrality as defined in Equation \eqref{betwenness}.

\begin{equation} \label{betwenness}
c_B(i) = \sum_{s \neq t \neq i \in V} \frac{\sigma(s, t \mid i)}{\sigma(s, t)},
\end{equation}

where \(\sigma(s, t \mid i)\) is the total number of shortest paths between nodes \(s\) and \(t\) that pass through node \(i\), and \(\sigma(s, t)\) is the total number of shortest paths between nodes \(s\) and \(t\), regardless of whether they pass through node \(i\) \cite{clark2018resilience}.
We note that while these centrality metrics provide valuable information regarding network topology, they are inherently limited by their dependency on connectivity patterns, without taking into consideration other critical factors such as economics, operational capacity, and system interdependencies that could influence node vulnerability in real-world scenarios. Therefore, we use external stressors (e.g., climate change-based heat extremes) as a critical factor that is independent of topology. Future research could also incorporate other factors and stressors, including recovery time estimates and cascading failure effects.  

\subsubsection{Disruption Scenarios}

This study examines five types of disruption scenarios: random node removal, targeted degree removal, targeted closeness removal, targeted betweenness removal, and targeted removal based on hot days.

Random node removal is a method where nodes in a network are randomly selected for removal. Targeted degree removal involves intentionally removing nodes with the highest degree (most connections) in a network to simulate and study the impact of losing critical nodes. Targeted closeness and targeted betweenness removal involve intentionally removing nodes that have the highest closeness or betweenness centrality, respectively. These methods help evaluate the network's efficiency and the role of key transit points or intermediaries. Targeted removal based on hot days involves identifying and prioritizing the removal of nodes (e.g., stations and ports) that are most affected by extreme heat waves, aiming to enhance the network's resilience against heat-related disruptions.

\subsubsection{Impact of Rising Hot Days} \label{Impact of Rising Hot Days}
As discussed in Section \ref{Literature Review}, one of the significant disruptions for infrastructure is heat waves. Critical temperature thresholds are essential in understanding the impact of extreme heat on infrastructure. Specific studies, such as \cite{bolitho2017heat}, identify 35$\si{\celsius}$ as a critical temperature above which infrastructure failures, such as rail track buckling and road pavement rutting, become more frequent.

For instance, in Phoenix, AZ, with average summer temperatures around 41$\si{\celsius}$ and highs reaching up to 50$\si{\celsius}$, infrastructure components like asphalt roads, steel bridges, and public transportation systems are significantly stressed. Projections suggest a sixfold increase in extreme heat events and a doubling of event duration by 2070. This underscores the necessity of adaptive strategies for infrastructure resilience to cope with the increasing frequency and intensity of heat waves, highlighting the urgency of this issue.

Based on the literature, we define hot days as the total number of days when temperatures exceed 35$\si{\celsius}$. To quantify projection‑driven uncertainty, we augmented the original single‑model analysis with an eight‑member, high‑resolution ensemble LOCA-2.

We used simulations of two experiments, ``historical'' and ``SSP5-8.5''.
In CMIP6 (Coupled Model Intercomparison Project Phase 6), the historical experiment ID is called ``historical''. 
This experiment is designed to simulate the past climate, specifically from the start of the industrial period, 1850, through the near-present, 2015. 
Shared Socioeconomic Pathway SSP5-8.5 (SSP5-8.5) experiment ID (for period 2015--2100) is referred to as fossil-fueled development and business-as-usual scenario \cite{Sharma_2023_Biogeosciences, Sharma_2022_JGRB}.
The ``8.5'' in SSP5-8.5 refers to the approximate level of radiative forcing (8.5 W/m²) by the year 2100. 
SSP5-8.5 assumes a world with a high dependency on fossil fuels like coal, oil, and natural gas; rapid economic development and a high population growth rate, especially in developing nations, drive demand for energy; and limited efforts are made to mitigate climate impacts or reduce emissions.

We analyse eight CMIP6 models, ACCESS-ESM1-5, CanESM5, GFDL-ESM4, INM-CM5-0, IPSL-CM6A-LR, MIROC6, MPI-ESM1-2-HR, and NorESM2-LM, under the high emissions SSP5-8.5 scenario for the 2021--2050 and 2051--2080 horizon and compare future scenarios with historical or baseline period of 1991--2020.
All these models are designed to simulate historical climate changes and variability, make centennial-scale projections of future climate, and generate initialized seasonal and decadal forecasts. It is an integral part of the Intergovernmental Panel on Climate Change assessments \cite{lee2023ipcc}.
This ensemble spans the accepted range of CMIP6 climate sensitivities and large‑scale circulation regimes while supplying the spatial granularity needed to resolve node‑scale hot‑day counts that coarse resolution Earth‑system models cannot capture. 
Using the 35$\si{\celsius}$ hot‑day threshold, we calculated the totals of hot days for 2021--2050 and 2051--2080 and at each freight node for each LOCA-2 model ensemble.

\subsection{Robustness and Resilience Metrics} 
In the context of a 
%intermodal
transportation system utilizing both rail and water modes, we conduct a robustness and resilience analysis. 
Our analysis assumes the availability of multiple rerouting options between node pairs. 

Evaluating the system's resilience necessitates assessing both the collapse and recovery processes. The initial step in this evaluation involves identifying appropriate metrics for critical functionality. 
In this study, we use the concept of the giant connected component (GCC)—~the largest interconnected group of nodes within a network~—to define Total Functionality (TF) as the number of nodes in the GCC when the network is fully operational. For our specific network, TF is determined to be 84. Fragmented Functionality (FF) is the number of nodes in the GCC at any given time after disruptions have caused the collapse of one or more nodes. The State of Critical Functionality (SCF) is a measure used to assess the operational state of a network following a disruption. We determine SCF for our network by calculating the ratio SCF = FF/TF, representing the proportion of the network's original connectivity that remains after the disruption \cite{clark2018resilience}.

\subsection{Potential Impact of Disruption on Freight Transportation}
The amount of tonnage transported via nodes can vary based on the location and size of the transport node and could be very different from the network topology.
To investigate how the disruptions in an unweighted and bi-directional network impact 
the total freight transported, we computed the total tonnage transportation capacity when the nodes were removed in the same order as in the robustness analysis.
This allows us to compare the impact of disruption of network topology on total freight transportation.
We used the tonnage projections for the year 2050 from the FTOT dataset for the impact assessment.

\section{Results and Discussion}
\label{RESULTS}

Here we present our results, focusing on the robustness and resilience of the transportation system under various disruption scenarios. The analysis is divided into several subsections, each addressing different aspects of system performance. First, we assess the system's resilience by evaluating its response to targeted and random disruptions, examining how they impact network connectivity and functionality. We then explore the effects of these disruptions on the freight transport capacity, providing insights into how the loss of critical nodes can significantly reduce the system's ability to transport resources. Additionally, we investigate the impact of rising temperatures on the network, highlighting the long-term implications of climate change on infrastructure. Finally, we rank the nodes with highest centrality measurements within the rail and water networks, identifying the most critical points that are pivotal in maintaining overall network resilience.

\subsection{Quantifying Increase in Hot Days Over Time}

\begin{figure}[!htbp]
    \centering
    \begin{subfigure}{0.9\columnwidth}
        \includegraphics[trim={0.1cm 1cm 0.0cm 0.9cm}, clip, width=\linewidth]{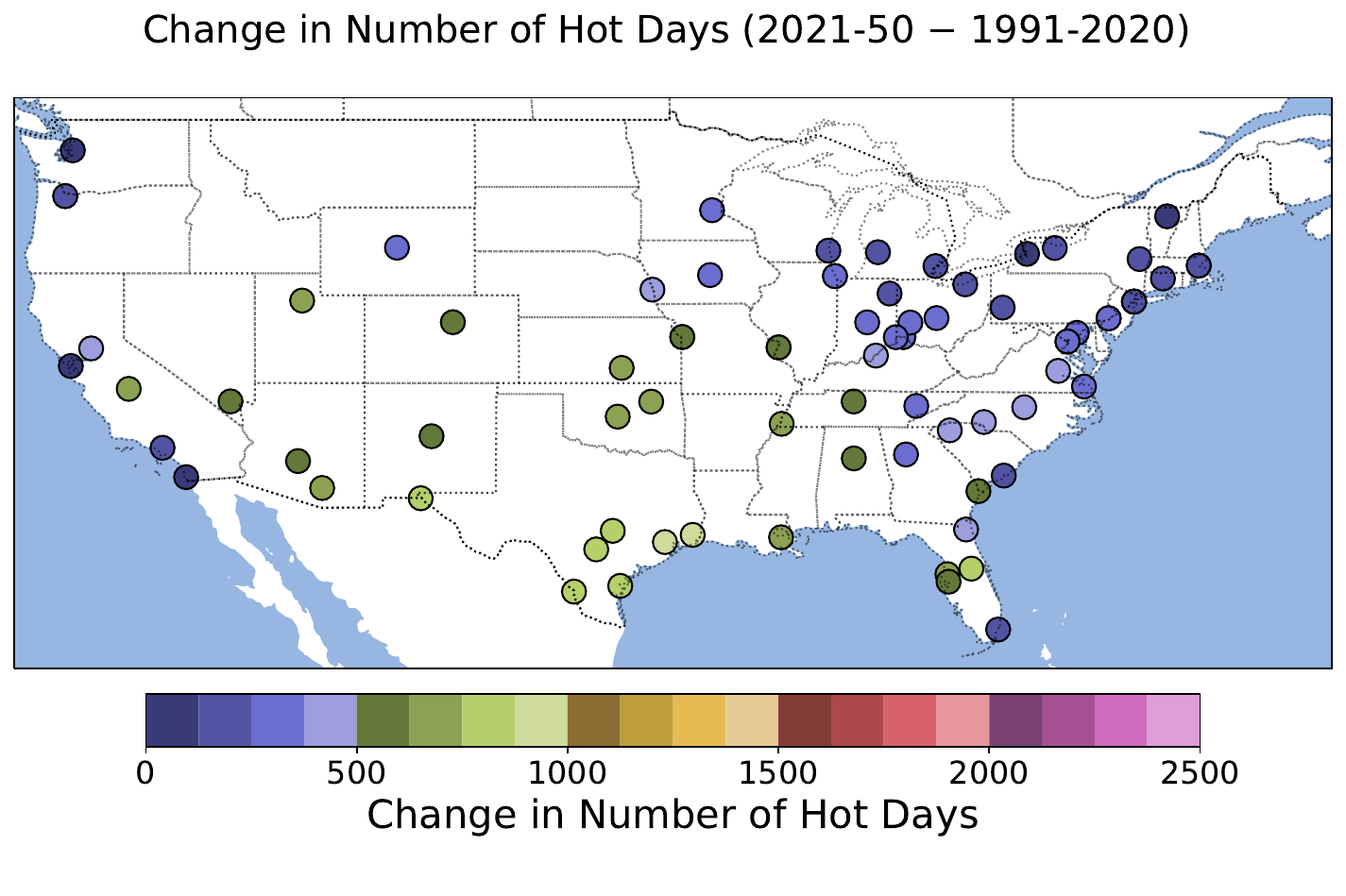}
        \caption{2021--2050 $-$ 1991--2020}
        \label{f:Change_HotDays_2021}
    \end{subfigure}
    
    \vspace{0.5cm}
    
    \begin{subfigure}{0.9\columnwidth}
        \includegraphics[trim={0.1cm 1cm 0.0cm 0.9cm}, clip, width=\linewidth]{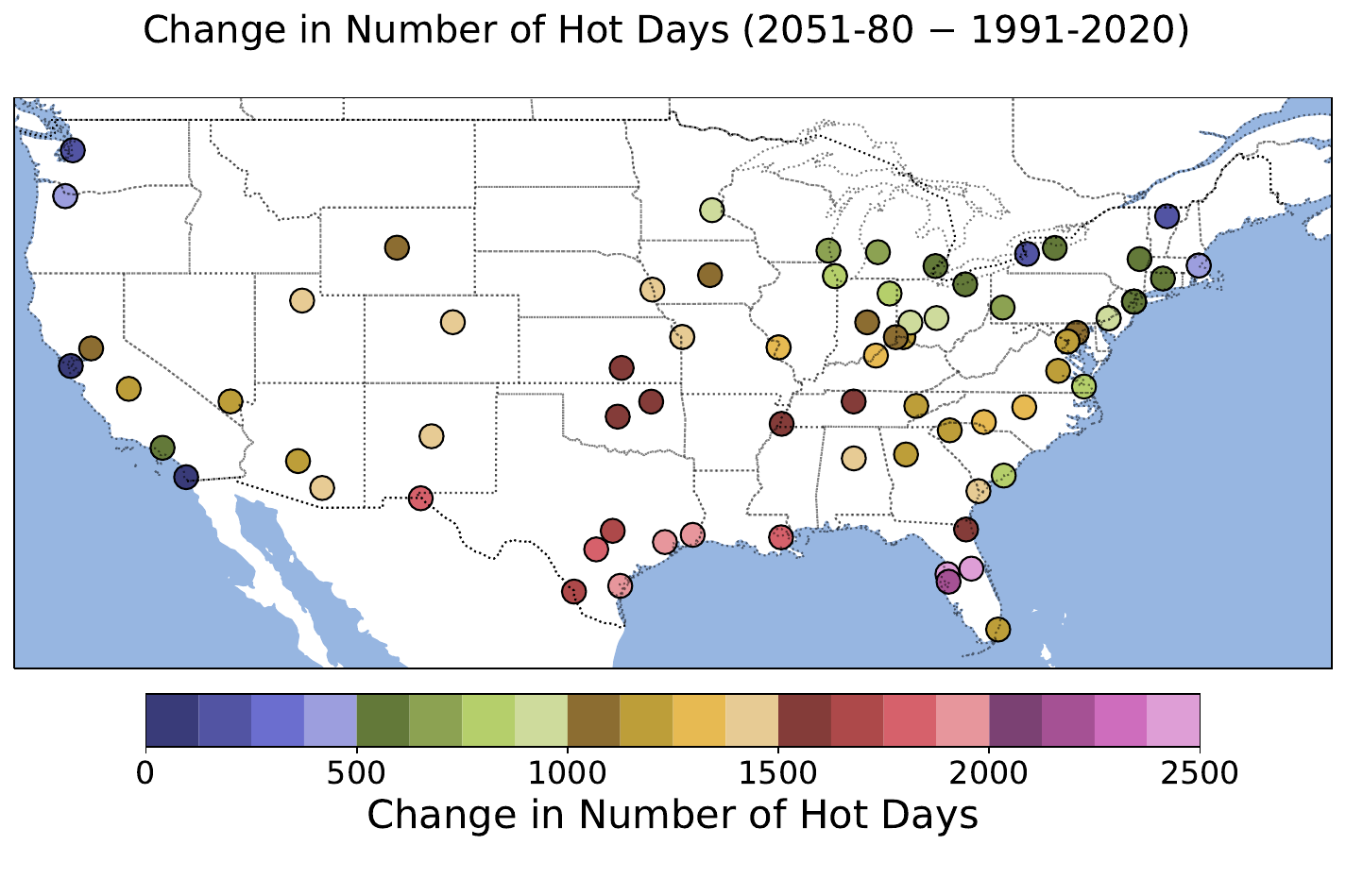}
        \caption{2051--2080 $-$ 1991--2020}
        \label{f:Change_HotDays_2051}
    \end{subfigure}

    \caption{\textbf{Change in Hot Days over Time}, averaged across eight CMIP6 models.
    Panel (a) shows the increase from 1991--2020 to 2021--2050.
    Panel (b) shows the increase from 1991--2020 to 2051--2080.
    The color bar range is kept constant for comparison.}
    \label{f:Heatwave_Conus}
\end{figure}

We computed the number of hot days for the historical period (1991--2020) and two future periods (2021--2050 and 2051--2080) of SSP5-8.5 scenario, as shown in (Figure~\ref{f:Heatwave_Conus}). Compared to 1991--2020, we see a significant increase in hot days during 2021--2050 (Figure~\ref{f:Heatwave_Conus}(a)), with Austin, TX, and San Antonio, TX, experiencing the largest rise, approximately 1,500 additional hot days.

Similarly, from 1991--2020 to 2051--- 2080 (Figure~\ref{f:Heatwave_Conus}(b)), Corpus Christi, TX, and Laredo, TX, show the highest increase of about 2,850 hot days.
 
The difference between the future and historical periods underscores the substantial increase in hot days and heat waves, which could severely impact infrastructure functionality over time. 

\subsection{Robustness and Resilience Analysis}
\begin{figure}[!htbp]
    \centering
    \begin{subfigure}{0.9\columnwidth}
        \includegraphics[trim={0.1cm 1cm 0.0cm 0.9cm}, clip, width=\linewidth]{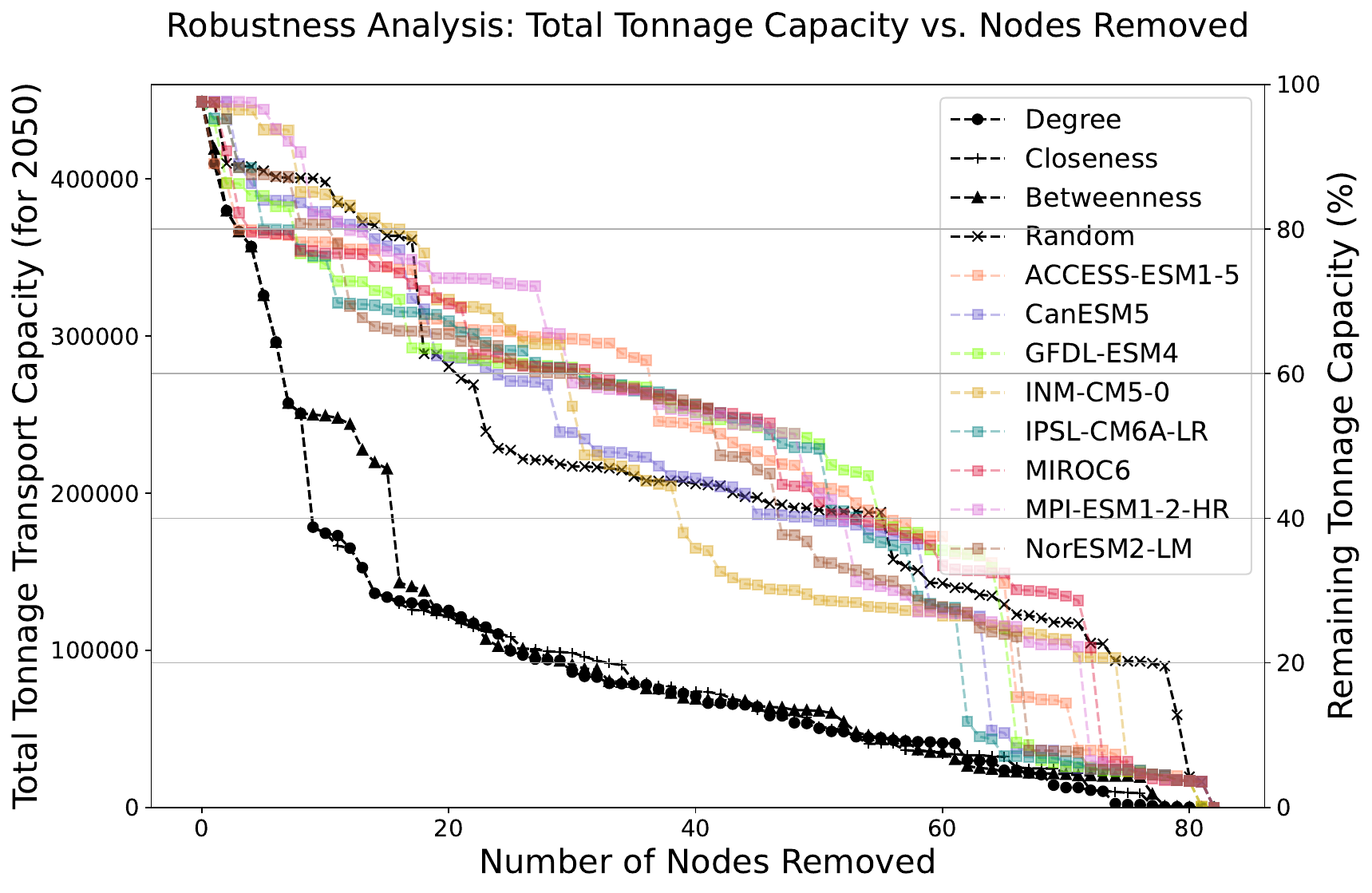}
        \caption{Rail Transport Robustness Analysis}
        \label{f:railfig}
    \end{subfigure}
    
    \vspace{0.5cm}
    
    \begin{subfigure}{0.9\columnwidth}
        \includegraphics[trim={0.1cm 1cm 0.0cm 0.9cm}, clip, width=\linewidth]{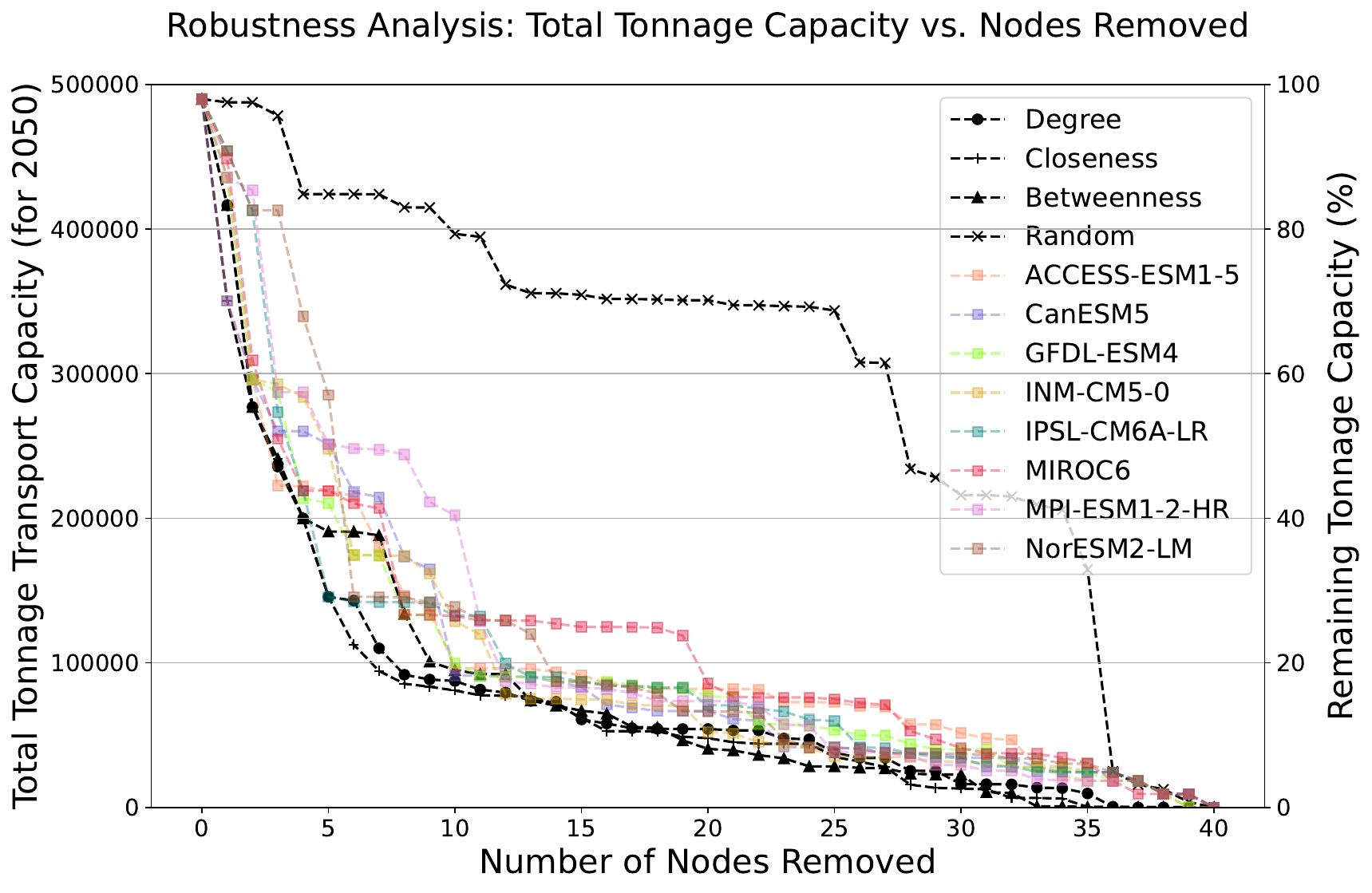}
        \caption{Water Transport Robustness Analysis}
        \label{f:waterfig}
    \end{subfigure}

    \caption{\textbf{Robustness Analysis of Rail and Water Freight Transport Networks.}
    The plots show the robustness analysis for five scenarios: Degree, Closeness, Betweenness Centrality, Random, and targeted attacks based on hot days from eight CMIP6 models.
    Panel (a) presents results for the unweighted, bidirectional rail network.
    Panel (b) presents results for the unweighted, bidirectional water network.}
    \label{f:robustnessFig}
\end{figure}

The resilience of the transportation system is assessed by evaluating its response to various disruption scenarios. Using the FTOT dataset, we analyzed the robustness of the rail and water networks, focusing primarily on visualizing the loss of critical functionality based on closeness, betweenness centrality, degree, climate change, and random disruption, as depicted in Figure~\ref{f:robustnessFig}.

TF was determined to be 84 nodes for the rail network and 47 for the water network. 
We quantified the robustness of our rail and water networks as they react to random and targeted disruptions. Our analysis revealed that the SCF, calculated as the ratio FF/TF, varied significantly depending on the type and severity of the disruptions. In scenarios where disruptions caused the collapse of several nodes, the SCF was notably reduced, indicating a significant impact on network connectivity and functionality. The network's ability to reroute and maintain operations was crucial in mitigating the effects of these disruptions.

We observe that node removal based on degree and dynamic centrality measures can lead to a more rapid network collapse. This is demonstrated by the targeted closeness, targeted betweenness, and targeted degree plots, which show a faster decline compared to the random and targeted hot days scenarios at Figure~\ref{f:robustnessFig}.

As shown in Figure~\ref{f:robustnessFig} (a), we assume that a loss of functionality up to 90$\%$ constitutes a complete collapse for our rail and water networks. For the rail network with a total of 84 nodes, complete collapse occurs when we lose 46$\%$ of nodes under the targeted degree scenario, 52$\%$ under targeted betweenness degree removal, 56$\%$ under targeted closeness degree removal%, 84$\%$ under targeted hot days removal,
and 87$\%$ under random removal.

Similarly, for the water network (see Figure~\ref{f:robustnessFig}(b)), a complete collapse happens with the loss of 23$\%$ of nodes under targeted degree removal, 32$\%$ under targeted betweenness degree removal, 40$\%$ under targeted closeness degree removal %79$\%$ under targeted hot days removal,
and 87$\%$ under random removal of nodes.

\subsubsection{Robustness Analysis under the LOCA‑Down‑scaled Climate‑Model Ensemble}
Across the eight LOCA-2 models, the qualitative ordering of disruption severity is unchanged.
For the rail network, collapse to an SCF $\leq$ 10\% occurs after 45\% ± 4\% of nodes are removed in the targeted sequence, with ACCESS‑ESM1‑5 yielding the earliest collapse (41\%) and GFDL‑ESM4 the latest (49\%) (Figure~\ref{f:robustnessFig}(a)).
The water network remains more brittle: fragmentation is reached after 24\% ± 3\% removals (Figure~\ref{f:robustnessFig}(b)).
Robustness of the rail and water network to hot days based disruption is much higher and comparable to random or stochastic disruptions.
The modest inter-model spread confirms that the robustness thresholds reported in the single-model baseline are not artifacts of a particular ESM or spatial resolution.

Ensemble‑frequency analysis 
%(Tables \ref{tab:rail_top10new} and \ref{tab:water_top10new}) 
shows that two rail hubs namely Houston, TX and Beaumont, TX enter the top‑10 hot‑day list in seven of eight climate models, while four water hubs (i.e., Houston TX, Beaumont TX, Corpus Christi TX, and Memphis TN) do so in all eight models. Thus, although down‑scaling sharpens spatial contrasts, it does not alter the identity of the locations that dominate heat‑related vulnerability; these hubs remain high‑priority candidates for heat‑resilient materials, power‑supply upgrades, and contingency‑routing protocols.

%Relative to the single‑model, 2.5°CanESM5 results reported earlier, the ensemble alters the headline metrics only marginally. 
For rail, the hot‑day collapse threshold shifts by at most ± 5 percentage points (46 \% in the coarse grid vs. 41–49\% across the ensemble); for water, the shift is ± 3 points (23\% vs. 21–26\%). Likewise, after 20 heat‑driven removals the residual rail tonnage averages 31\% ± 3\%, bracketing the 30\% value obtained with the coarse model. These small deviations demonstrate that our core findings—rapid capacity loss under heat‑related outages and the outsized importance of Gulf‑coast hubs—are robust to both spatial resolution and climate‑model choice, providing added confidence for policy application.

Figures~\ref{f:robustnessFig} and ~\ref{f:tonnageloss} condense the eight hot‑day curves for rail and water onto single panels alongside the unchanged degree‑, closeness‑, betweenness‑targeted, and random baselines. By viewing all models together, the relative magnitudes of climate uncertainty and structural uncertainty become apparent: the coloured hot‑day curves occupy a narrow corridor, whereas the gap between any hot‑day curve and its corresponding centrality curve is much larger. This visual synthesis reinforces three practical conclusions.

First, climate‑model spread is smaller than structural spread. The inter‑model fan is modest compared with the gap between hot‑day and centrality‑targeted sequences, indicating that network structure rather than exact hot‑day count dominates resilience outcomes. Second, policy‑critical thresholds are stable. Whether one uses a coarse resolution ESM or a 6 km ensemble, the number of heat‑vulnerable nodes whose failure precipitates systemic collapse shifts by no more than ± one or two nodes for rail and ± one node for water. Lastly, capacity erodes far more quickly than connectivity. Even under the most lenient hot‑day scenario, system‑wide freight tonnage falls below 35\% once approximately twenty rail nodes—or just five major water nodes—have been removed. This stark disparity underscores the operational importance of evaluating tonnage capacity in addition to topological metrics.

\subsection{Impact on Freight Transport Capacity}

\begin{figure}[!htbp]
    \centering
    \begin{subfigure}{0.9\columnwidth}
        \includegraphics[trim={0.1cm 1cm 0.0cm 0.9cm}, clip, width=\linewidth]{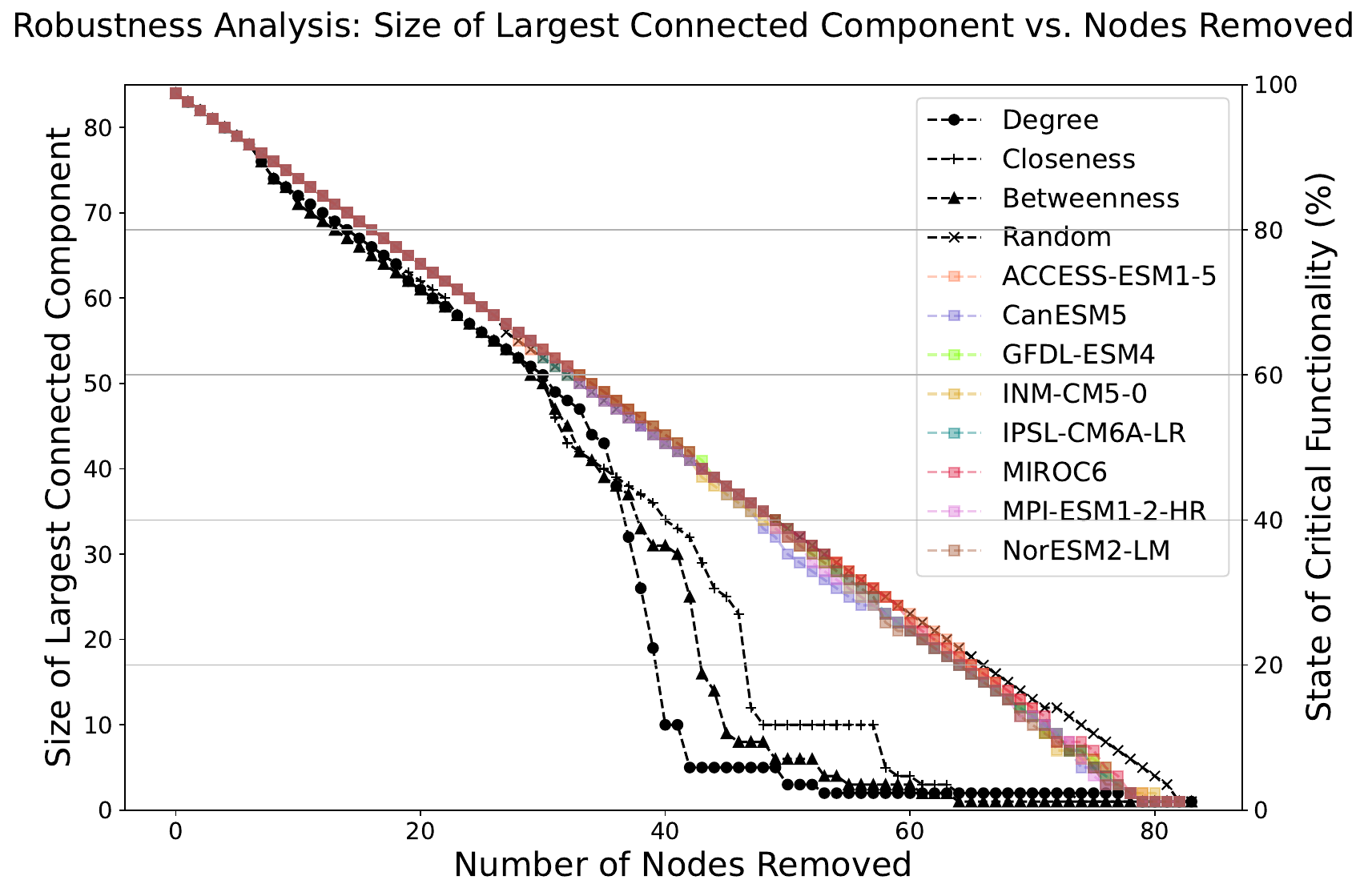}
        \caption{Rail Transport Analysis}
        \label{f:railfigton}
    \end{subfigure}
    
    \vspace{0.5cm}
    
    \begin{subfigure}{0.9\columnwidth}
        \includegraphics[trim={0.1cm 1cm 0.0cm 0.9cm}, clip, width=\linewidth]{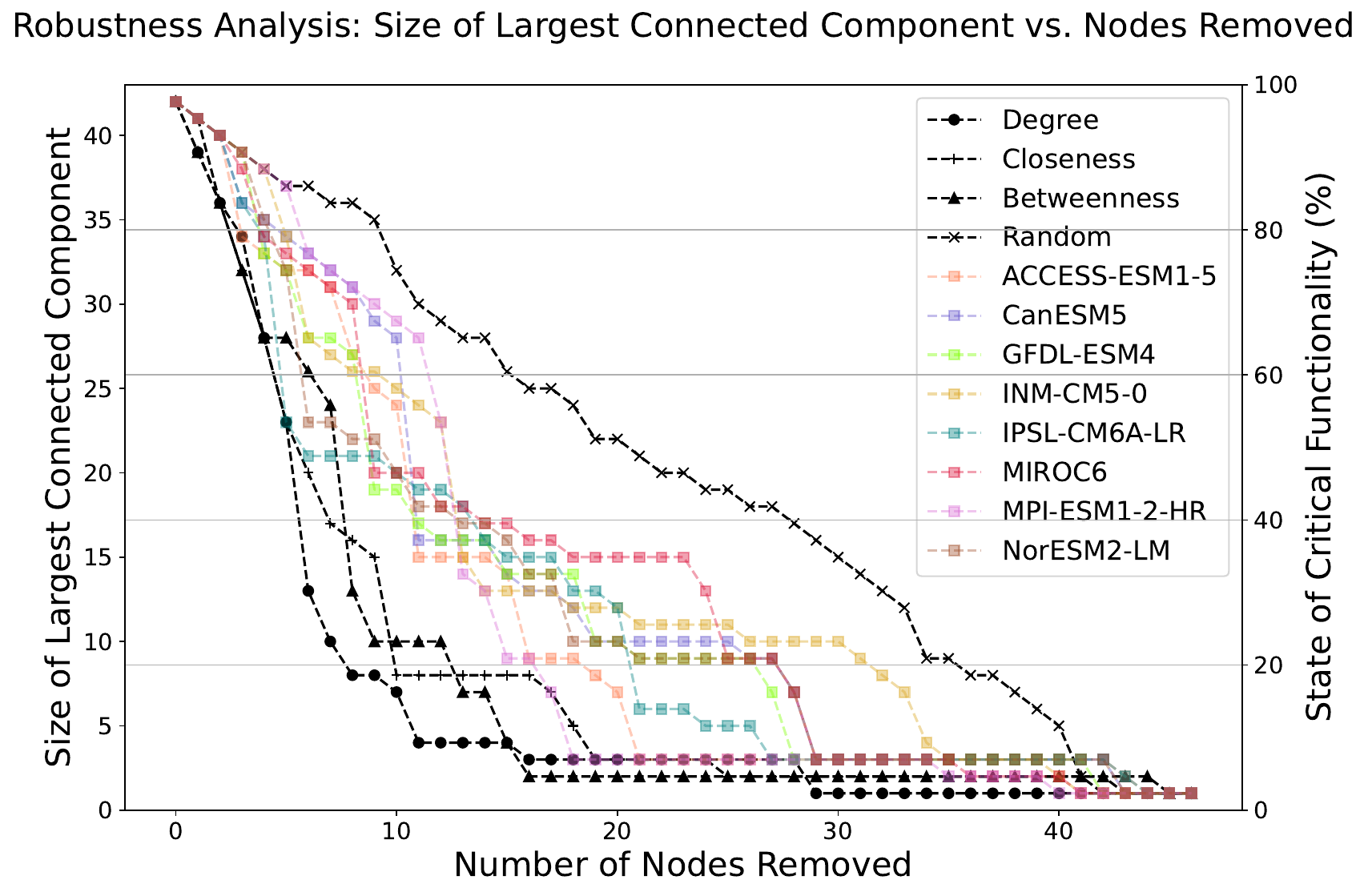}
        \caption{Water Transport Analysis}
        \label{f:waterfigton}
    \end{subfigure}

    \caption{\textbf{Potential Loss of Load Transport Capacity of Rail and Water Modes.}
    Panel (a) shows the impact of node removal on the total tonnage capacity of the rail freight network.
    Panel (b) shows the same for the water freight network.}
    \label{f:tonnageloss}
\end{figure}

The impact of disruption on the load-carrying capacity of the network under various
disruptions is shown in Figure~\ref{f:tonnageloss}.
To allow a comparison of the robustness of network topology with the total tonnage capacity of the network, the nodes in in Figure~\ref{f:tonnageloss} are removed in the same order as shown in Figure~\ref{f:robustnessFig}.
Since the distribution of tonnage varies across nodes, the robustness of total tonnage capacity (see Figure~\ref{f:tonnageloss}) could be different from the robustness of unweighted network topology (see Figure~\ref{f:robustnessFig}). 
While the robustness analysis indicated that the potential impact of rising hot days on
the rail network was minimal (Figure~\ref{f:robustnessFig}(a)), the impact on the 
total tonnage transported via rail is very large (Figure~\ref{f:tonnageloss}(a)).

The impact on rail freight transport carrying capacity (Figure~\ref{f:tonnageloss}(a)) under targeted removal of 
nodes -~based on degree, closeness, and betweenness centrality~- was higher than 
the SCF under the same scenarios (Figure~\ref{f:robustnessFig}(a)). 
This highlights that targeted disruption of only 10 and 20 nodes can hinder the 
rail freight transport and reduce it to only about 40$\%$ and 20$\%$ of total tonnage potential.
The exception to this pattern was the random removal of nodes, where the impact on freight-carrying capacity was less than the SCF. 

Similarly, impact on water freight transport carrying capacity (Figure~\ref{f:tonnageloss}(b)) under targeted removal of 
nodes -~based on degree, closeness, betweenness centrality, and hot days~- was higher than the respective SCF under the same scenarios (Figure~\ref{f:robustnessFig}(b)). 

These results highlight the importance of investigating both the network topology and impact metrics to measure the holistic impact of climate change or targeted disruptions on the system's robustness. 
Removal of only a few critical nodes can drastically reduce the ability of the total system to fulfill its objective of transporting resources on time. Future work could refine the analysis with flow‑weighted centralities such as traffic betweenness \cite{kazerani2009can} or current‑flow closeness \cite{brandes2005centrality} once higher‑resolution origin–destination data becomes available, thereby reducing the bias of topology‑only metrics.

\subsection{Rising Temperature Impact On Network Topology}

The analysis of temperature data from the eight LOCA‑down‑scaled climate models revealed a substantial increase in the number of hot days over the periods of 2021-2050 and 2051-2080 compared to the historical period of 1991-2020 (see Figure~\ref{f:Heatwave_Conus}).

\subsection{Node, Centrality Measurements and Hot Days}

\begin{table}[!ht]
    \caption{\textbf{Top 10 Rail Freight Transport Nodes Removed. Border nodes' names are combination of city and state name in FTOT dataset.}}
    \label{tab:top10}
    \begin{center}
        \begin{small}
        %\resizebox{\textwidth}{!}{%
        \begin{tabular}{l l l l}
            Rank & Degree & Betweenness & Closeness \\\hline
            1  & Houston          & Houston          & Chicago IL                 \\
            2  & Chicago IL       & Chicago IL       & Los Angeles                \\
            3  & Iowa             & Iowa             & Atlanta                    \\
            4  & Los Angeles      & Los Angeles      & Houston                    \\
            5  & Baton Rouge      & Atlanta          & Cleveland                  \\
            6  & New Orleans LA   & New Orleans LA   & New York NY                \\
            7  & Atlanta          & Baton Rouge      & Philadelphia PA            \\
            8  & Wyoming          & Wyoming          & Iowa                       \\
            9  & Detroit          & Cleveland        & San Francisco              \\
            10 & Fort Wayne       & San Francisco    & Minneapolis–St.\ Paul MN   \\\hline
        \end{tabular}
        %}
        \end{small}
    \end{center}
\end{table}
\begin{table}[!ht]
    \caption{\textbf{Top 10 Water Freight Transport Nodes Removed.} Border node names are a combination of city and state names from the FTOT dataset.}
    \label{tab:top10water}
    \begin{center}
        \begin{small}
        \begin{tabular}{l l l l}
            \hline
            Rank & Degree & Betweenness & Closeness \\
            \hline
            1  & New Orleans LA & New Orleans LA & New Orleans LA \\
            2  & Baton Rouge & Houston & Houston \\
            3  & Houston & Baton Rouge & Baton Rouge \\
            4  & Corpus Christi & Corpus Christi & St. Louis IL \\
            5  & Beaumont & St. Louis IL & San Francisco \\
            6  & Chicago IL & Cincinnati KY & Los Angeles \\
            7  & St. Louis IL & Philadelphia DE & Chicago IL \\
            8  & Pittsburgh PA & Pittsburgh PA & Memphis TN \\
            9  & Cincinnati KY & Seattle & St. Louis MO \\
            10 & St. Louis MO & Beaumont & Cincinnati KY \\
            \hline
        \end{tabular}
        \end{small}
    \end{center}
\end{table}

Table~\ref{tab:top10} and Table~\ref{tab:top10water} highlight the top 10 nodes in the rail and water networks separately, ranked by their centrality measures: degree, betweenness, and closeness. These centrality metrics are crucial in identifying the most critical nodes significantly influencing overall network connectivity and functionality. In the rail network, Houston, TX, and Chicago, IL-IN-WI (IL Part), emerge as the most critical nodes across multiple centrality measures. This indicates that these locations are not only heavily connected (degree) but also play crucial roles in facilitating the flow of freight across the network (betweenness) and are well-positioned relative to other nodes (closeness). The high centrality scores of these nodes suggest that disruptions here could have widespread effects, potentially leading to substantial declines in network efficiency and resilience. Other notable nodes include Iowa and Los Angeles, CA, which rank highly in degree and betweenness, further emphasizing their importance in the rail network.
For the water network, New Orleans, LA-MS (LA Part), stands out as the most critical node across all three centrality measures, underscoring its pivotal role in maintaining the connectivity and flow of goods in the network. Baton Rouge, LA, and Houston, TX, also rank highly, particularly in degree and betweenness, suggesting they are crucial hubs for water-based freight transport. The prominence of these nodes indicates that they are essential for ensuring the robustness and resilience of the water network.

\section{Conclusion}
\label{Conclusion}

We provide an in-depth examination of the robustness and resilience of rail and water transportation networks under different disruption scenarios.
By examining the dynamic response of the transportation system to different types of disruptions, we conduct a thorough evaluation of the robustness and resiliency of the rail and water freight transportation system of the US. 
Measuring robustness is a critical initial step toward improving the system's resilience. 
This paper extends the robustness analysis by investigating the potential impact of climate change on the network topology and the total load capacity of the network.

 It is noteworthy that heat wave disruptions predominantly target cities in Texas, which have high connectivity (degree). 
 These cities are also likely to face disruptions under targeted degree removal scenarios. Stakeholders and policymakers should recognize that increasing temperatures could pose significant challenges to the freight transportation system in certain regions and should consider this in future resilience planning.

This study adopts the largest connected component and SCF indicators because they are transparent, data‑lean, and widely used for nationwide screening. However, these topology‑based measures do not capture the adaptive behaviours that unfold once a disruption occurs—such as dynamic rerouting, queue formation, surge capacity, or carrier‑specific priority rules. A natural extension is to embed the rail‑and‑water network in an agent‑based or system‑dynamics framework capable of modelling time‑dependent logistics: trains or barges joining queues, temporary speed restrictions, pop‑up container depots, and contract‑driven diversion choices. Developing and calibrating such a behaviourally rich model lies beyond the scope of the present paper, but we flag it as a priority for future work that will translate these first‑order vulnerability screens into detailed operational playbooks.

The methods described in this paper can assist policymakers, including Departments of Transportation (DOTs) and Metropolitan Planning Organizations (MPOs), in quantifying the impact of disruptions under various scenarios at smaller scales.

There are some uncertainties and practical challenges: the model‐based insights presented here must be interpreted in light of three uncertainty layers. (i) Climate‐model spread: addressed via the LOCA-2 eight‐model ensemble of CMIP6 high resolution downscaled models. (ii) Socio economic pathway uncertainty affects freight demand trajectories; the SSP5‑8.5 tonnage projections used here represent an upper‑bound stress test. (iii) FTOT economic assumptions underlie the tonnage weights, meaning commodity‑specific or value‑at‑risk adjustments are advisable before final investment decisions. Translating model outputs into action also faces practical constraints: finite maintenance windows, multi‑agency permitting, and limited diversion routes can delay the rerouting steps we simulate. Federal Railroad Administration guidance on phased speed‑restriction implementation during heat waves illustrates how analytical thresholds must be coordinated with operational realities. Despite these caveats, the analysis delivers three actionable take‑aways for practitioners:
First, a set of ensemble‑robust heat‑vulnerability hotspots, Houston TX, Beaumont TX, Corpus Christi TX, Memphis TN and New Orleans LA, emerges consistently across all climate‑model runs, indicating where heat‑resilient track materials and power‑supply upgrades should be prioritised. Second, the results establish early‑warning thresholds: once roughly twenty rail hubs or five major ports are lost, system‑wide freight‑tonnage capacity drops below 35\%, a level that should trigger contingency‑routing plans. Third, because freight capacity deteriorates far more rapidly than network connectivity, limited resilience budgets are best allocated to these high‑throughput nodes first, while lower‑volume yet topologically redundant facilities can be scheduled for subsequent investment phases.

To facilitate practical application, DOTs and MPOs could adopt our resilience analysis framework to assess the robustness of their specific regional transportation networks. By tailoring the network analysis to local contexts, these organizations might identify critical nodes and links within their jurisdictions that are most vulnerable to disruptions, including those driven by climate change. Implementing this process involves collecting and analyzing localized data on transportation infrastructure, freight flows, and potential climate impacts, while ensuring that operational security is not compromised. Agencies should implement data management protocols to protect sensitive information, such as data anonymization and controlled access, to prevent exposure of vulnerabilities to nefarious actors.
%%%
 
%%%
The findings emphasize the importance of identifying critical nodes and analyzing temperature impacts. To address the growing challenges posed by climate change and extreme weather events, stakeholders must prioritize resilience planning and invest in infrastructure adaptations. Collaborating with organizations such as the American Association of State Highway and Transportation Officials (AASHTO) can facilitate the integration of our resilience framework into state-level practice, promoting standardization and knowledge sharing among different jurisdictions. 
Although FTOT currently provides the most advanced, nationally consistent multimodal freight data set available to the public, its centroid representation and macro‑economic assumptions remain sources of uncertainty. Should higher‑resolution or independently compiled national freight data sets become available, we would integrate them in vulnerability thresholds in future studies. Future research should focus on high-resolution models and multi-threaded analyses to further refine resilience strategies and ensure the sustained operational efficiency of transportation networks.

In future studies, we plan to incorporate the road mode into our model, measuring the resiliency and robustness of road nodes in the same manner as we did for the rail and water networks. 
Additionally, we aim to integrate intermodality in transportation by including road, rail, and water modes within our transportation system to evaluate the benefits and challenges of intermodal transportation. 
Future studies could include other climate extremes and compound extremes that are likely to increase over time, such as floods and hurricanes, and have a larger impact on stressed systems.
Furthermore, we will investigate various recovery strategies to identify the most effective methods for restoring normal functionality following disruptive events. This will help us develop comprehensive resilience plans that ensure the transportation system can quickly recover and maintain operational stability.

\section{acknowledgement}
\label{acknowledgement}

This research was conducted at the University of Tennessee, Knoxville, in cooperation with the Oak Ridge National Laboratory. This work was supported in part by the US Department of Energy’s Advanced Research Projects Agency-Energy (ARPA-E) under the project (\#DE-AR0001780), titled ``A Cognitive Freight Transportation Digital Twin for Resiliency and Emission Control Through Optimizing Intermodal Logistics'' (RECOIL). This manuscript has been authored by UT-Battelle, LLC, under contract DE-AC05-00OR22725 with the US Department of Energy (DOE). The US government retains and the publisher, by accepting the article for publication, acknowledges that the US government retains a nonexclusive, paid-up, irrevocable, worldwide license to publish or reproduce the published form of this manuscript, or allow others to do so, for US government purposes. DOE will provide public access to these results of federally sponsored research in accordance with the DOE Public Access Plan (http://energy.gov/downloads/doe-public-access-plan).

\section{author contributions}
\label{author_contributions}

The authors confirm their contribution to the paper as follows: \textit{Study Conception and Design:} all authors (Maedeh Rahimitouranposhti, Bharat Sharma, Mustafa Can Camur, Olufemi A. Omitaomu, Xueping Li); \textit{Data Collection:} all authors; \textit{Analysis and Interpretation of Results:} all authors; \textit{Draft Manuscript Preparation:} all authors. All authors review the results and approve the final version of the manuscript.

\section{Conflict of Interest}
The authors declared no potential conflicts of interest with respect to the research, authorship, and/or publication of this article.

% \newpage
\bibliographystyle{trb}
 \bibliography{bib_file}

\end{document}